\documentclass{esannV2}
\usepackage[dvips]{graphicx}
\usepackage[utf8]{inputenc} 
\usepackage{amssymb,amsmath,array}
\usepackage{bbold}
\usepackage{physics}
\usepackage{xcolor}
\usepackage{hyperref}
\usepackage{tikz}
\usepackage{ifthen,xstring}
\usetikzlibrary{calc}
\usetikzlibrary{external}
\usepackage{bbold}
\usepackage[justification=centering]{caption}
\DeclareMathOperator*{\argmin}{arg\,min}

\usepackage{todonotes}
\usepackage{ulem}
\begin{document}

\definecolor{lblue}{rgb}{.1,.6,1}
\colorlet{dyell}{yellow!75!black}
\colorlet{dgreen}{green!50!black}
\colorlet{orng}{yellow!50!red}
\def\isocolor{BurntOrange}
\def\unicolor{Aquamarine}
\def\nodecolor{LimeGreen}

\definecolor{norm}{HTML}{79ADDC}
\definecolor{conj}{HTML}{ffc09f}
\definecolor{data}{HTML}{ffee93}
\definecolor{proj}{HTML}{ffe247}
\definecolor{colora}{HTML}{eba7b1}
\definecolor{colorb}{HTML}{a7d3eb}
\def\norm{norm}
\def\conj{conj}
\def\data{data}
\def\proj{proj}
\def\colora{colora}
\def\colorb{colorb}

\newcommand{\TNode}[4]{ 
	\coordinate (temp) at (#2);
	\ifthenelse{\equal{#1}{}}{
		\draw[line width=1pt,fill=#3] (temp)circle(.25);}
	{\ifthenelse{\equal{#1}{C}}{
			\draw[rounded corners=2pt,line width=1pt,fill=#3] ($(temp)-(.25,.25)$)rectangle++(.5,.5);}
		{\ifthenelse{\isodd{#1}}{
				\draw[rounded corners=2pt,line width=1pt,fill=#3] (temp)--++($(.25,0)+#1*(0,.25)$)--++($#1*(0,-.5)$)--++(-.5,0)--(temp);
				\coordinate (temp) at ($(temp)-#1*(0,.05)-(.05,0)$);}
			{\draw[rounded corners=2pt,line width=1pt,fill=#3] (temp)--++($(-.25,0)+#1*(0,.125)$)--++($#1*(0,-.25)$)--++(.5,0)--(temp);
				\coordinate (temp) at ($(temp)-#1*(0,.05)-(.05,0)$);}}}
	\node at (temp) {#4}}

\newcommand{\Gate}[5]{ 
	\coordinate (temp) at (#2);
	\ifthenelse{\equal{#1}{V}}{
		\draw[rounded corners=2pt,line width=1pt,fill=#4!50!white] ($(temp)-(.25,.25)$)rectangle++($(-.25,.5)+#3*(.75,0)$)node[midway]{#5}}
	{\draw[rounded corners=2pt,line width=1pt,fill=#4!50!white] ($(temp)-(.25,-.25)$)rectangle++($(.5,.25)-#3*(0,.75)$)node[midway]{#5}}}

\newcommand{\Qin}[2]{ 
	\coordinate (temp) at (#2);
	\StrLen {#1}[\tmpLn]
	\ifthenelse{ \tmpLn= 1}{
		\ifthenelse{\equal{#1}{V}}{
			\draw[line width=1pt, fill=yellow!75!red] ($(temp)-(0,.1)$)circle(.15)}
		{\draw[line width=1pt, fill=yellow!75!red] ($(temp)-(.1,0)$)circle(.15)}}
	{\ifthenelse{\equal{#1}{V0}}{
			\node[below] at (temp) {$|0\rangle$}}
		{\node[left,align=right] at (temp){$|0\rangle$}}}}

\newcommand{\Qout}[3]{ 
	\coordinate (temp) at (#2);
	\StrLen {#1}[\tmpLn]
	\ifthenelse{ \tmpLn= 1}{
		\ifthenelse{\equal{#1}{V}}{
			\draw[line width=1pt] ($(temp)+(-.1,.05)$) --++(.2,.1)node[above]{#3};
			\draw[line width=1pt] ($(temp)+(-.1,0)$) --++(.2,.1)}
		{\draw[line width=1pt] ($(temp)+(-.05,.1)$) --++(.1,-.2)node[right]{#3};
			\draw[line width=1pt] ($(temp)+(0,.1)$) --++(.1,-.2)}}{
		\ifthenelse{\equal{#1}{VM}}{\coordinate (temp2) at ($(temp)+(0,.25)$);}{\coordinate (temp2) at ($(temp)+(.25,0)$);}
		\draw[line width=1pt] ($(temp2)-(.25,.25)$)rectangle++(.5,.5);\draw[line width=1pt,->] ($(temp2)+(-.25,-.25)$)--++(.35,.35);}}

\newcommand{\Qreset}[3]{
	\coordinate (temp) at (#2);
	\ifthenelse{\equal{#1}{V}}{
		\ifthenelse{\equal{#3}{in}}{
			\draw[fill=white,color=white] ($(temp)-(.1,-.1)$)rectangle++(.2,.4);
			\Qout{V}{$(temp)$}{};
			\Qin{V}{$(temp)+(0,.5)$}}{
			\ifthenelse{\equal{#3}{out}}{\draw[fill=white,color=white] ($(temp)-(.1,-.1)$)rectangle++(.2,1.2);
				\Qout{VM}{$(temp)$}{};
				\Qin{V0}{$(temp)+(0,1.2)$}}{\draw[fill=white,color=white] ($(temp)-(.1,-.1)$)rectangle++(.2,.4);
				\Qout{V}{$(temp)$}{};
				\Qin{V0}{$(temp)+(0,.5)$}}}}{
		\ifthenelse{\equal{#3}{in}}{
			\draw[fill=white,color=white] ($(temp)-(0,.1)$)rectangle++(.4,.2);
			\Qout{H}{$(temp)$}{};
			\Qin{H}{$(temp)+(.5,0)$}}{
			\ifthenelse{\equal{#3}{out}}{
				\draw[fill=white,color=white] ($(temp)-(0,.1)$)rectangle++(1.1,.2);
				\Qout{HM}{$(temp)$}{};
				\Qin{H0}{$(temp)+(1.2,0)$}}{\draw[fill=white,color=white] ($(temp)-(0,.1)$)rectangle++(.65,.2);
				\Qout{H}{$(temp)$}{};
				\Qin{H0}{$(temp)+(.75,0)$}}}}}

\def\ctrlsize{0.6pt}

\newcommand{\MPSiso}[4]{
	\coordinate (I1) at (#1);
	\ifthenelse{\equal{#4}{d}}
	{
		\ifthenelse{\equal{#3}{r}}
		{   
			\draw[rounded corners=2pt,line width=1pt,fill=#2]  
			(I1)--++(.25,.25)--++(-.5,0)--++(0,-.5)--(I1);}
		{
			\draw[rounded corners=2pt,line width=1pt,fill=#2] 
			(I1)--++(-.25,.25)--++(.5,0)--++(0,-.5)--(I1);}}
	{   
		\ifthenelse{\equal{#3}{r}}
		{
			\draw[rounded corners=2pt,line width=1pt,fill=#2] 
			(I1)--++(-.25,.25)--++(0,-.5)--++(.5,0)--(I1);}
		{   
			\draw[rounded corners=2pt,line width=1pt,fill=#2] 
			(I1)--++(.25,.25)--++(0,-.5)--++(-.5,0)--(I1);}}        
}

\newcommand{\Projector}[2]{
	\coordinate (I1) at (#1);
	\draw[rounded corners=2pt,line width=1pt,fill=#2]  
	(I1)--++(.25,0)--++(0,-.25)--++(-.5,0)--++(0,.25)--(I1);
}

\newcommand{\symiso}[2]{
	\coordinate (I1) at (#1);  
	\draw[line width=1pt,fill=#2] 
	(I1)--++(0.35,0)--++(-0.35,0.35)--++(-0.35,-0.35)--(I1);
}

\newcommand{\symuni}[2]{
	\coordinate (I1) at (#1);
	\draw[line width=1pt,fill=#2] 
	($(I1)$)--++(0.25,0)--++(0,0.5)--++(-.5,0)--++(0,-.5)--($(I1)$);
}

\newcommand{\symnode}[2]{
	\coordinate (I1) at ($(#1)$);
	\draw[line width = 1pt, fill=#2] (I1) circle (.25);
}

\newcommand{\Cnot}[2]{
	\def\circlesize{0.06}
	\coordinate (temp) at (#1);
	\draw[line width = 1pt] (temp)--++($(0,#2)$);
	\draw[fill=black] (temp) circle (\ctrlsize);
	\draw[fill=white,line width = 1pt] ($(temp)+(0,#2)$) circle (\circlesize);
	\draw[line width = 1pt] ($(temp)-(\circlesize,-#2)$) --++ ($(2*\circlesize,0)$)
	\draw[line width = 1pt] ($(temp)-(0,-#2+\circlesize)$) --++ ($(0,2*\circlesize)$)
}

\newcommand{\Cphase}[2]{
	\coordinate (temp) at (#1);
	\draw[line width = 1pt] (temp)--++($(0,#2)$);
	\draw[fill=black] (temp) circle (\ctrlsize);
	\draw[fill=black] ($(temp)+(0,#2)$) circle (\ctrlsize);
}

\newcommand{\Cunitary}[3]{
	\coordinate (temp) at (#1);
	\draw[line width = 1pt] (temp)--++($(0,#2)$);
	\draw[fill=black] (temp) circle (\ctrlsize);
	\draw[line width = 1pt, fill = white] ($(temp)+(-0.1,-0.1+#2)$) rectangle ++(0.2,0.2);
	\node[line width = 1pt] at ($(temp)+(0,#2)$) {#3};
}

\newcommand{\Hadamard}[1]{
	\coordinate (temp) at (#1);
	\draw[line width = 1pt, fill = white] ($(temp)+(-0.1,-0.1)$) rectangle ++(0.2,0.2);
	\node[line width = 1pt] at (temp) {H};
}
\newcommand{\unitaryone}[2]{
	\coordinate (temp) at (#1);
	\draw[line width = 1pt, fill = white] ($(temp)+(-0.1,-0.1)$) rectangle ++(0.2,0.2);
	\node[line width = 1pt] at (temp) {#2};
}
\newcommand{\unitarytwo}[3]{
	\coordinate (temp) at (#1);
	\draw[line width = 1pt, fill = white] ($(temp)+(-0.1,-0.1)$) rectangle ++($(0.2,0.2+#2)$);
	\node[line width = 1pt] at ($(temp)+(0,#2*0.5)$) {#3};
} 

\title{Quantum Tensor Network Learning with DMRG}
\author{Gustav J L Jäger$^{1,2}$ Martin B Plenio$^2$ Hans-Martin Rieser$^1$
%
%
\vspace{.3cm}\\
%
1- Deutsches Zentrum für Luft- und Raumfahrt e.V. -- Institut für KI-Sicherheit\\
Wilhelm-Runge-Straße 10, 89081 Ulm, Germany 
%
\vspace{.1cm}\\
2- Universität Ulm -- Institut für Theoretische Physik and IQST,\\ Albert-Einstein-Allee 11, 89081 Ulm, Germany\\
}

\maketitle

\begin{abstract}
Tensor Networks are a relatively new machine learning approach. The architectures proposed initially are inspired by approaches from quantum many-body physics simulations. One common layout is the matrix product state (MPS) also known as a tensor train optimized with gradient descent techniques. We introduce a global normalization condition, so that the MPS represents a quantum state. We investigate two optimization methods that find the locally optimal tensors and compare them regarding their effectiveness. One is based on gradient descent and the other on an adaptation of DMRG.
\end{abstract}

\section{Introduction}
Tensor networks (TN) originate from the field of quantum many-body physics \cite{Orus2013,Bridgeman2017}. They are used as a prominent method to simulate quantum systems \cite{Ranbook2020}, and approximate their properties such as ground states \cite{Schollwöck2005}. Their main advantage lies in their ability to approximate large quantum states with limited storage. Quantum-inspired tensor networks have found an application in classical machine learning models, see \cite{Stoudenmire2016}.
\par
The application of tensor networks to quantum machine learning connects both fields \cite{Rieser2023}. To raise the whole potential of TN approaches developed in quantum physics it is necessary to transfer the local optimization techniques, namely Density Matrix Renormalization Group (DMRG), to the machine learning context. In this work we explore how this DMRG approach with the underlying Lanczos method to quantum machine learning. One challenge that arises, is that quantum states and quantum channels have normalization conditions due to quantum probability conservation \cite{Wilde2017}.
\par
The basic tensor network ansatz we investigate is inspired by \cite{Stoudenmire2016}. However, we introduce a normalization condition on the tensor network, such that contracting it yields a normalized vector that may be mapped to a quantum state. We then present how the gradient descent algorithm can be modified to take into account this condition and how a modified DMRG algorithm can be applied for optimization as well. The latter is based on a post processing step taking into account the loss function of the machine learning problem. Finally, we compare how the introduction of the normalization condition impacts the accuracy and the loss, when classifying the MNIST dataset.
\section{Tensor Network based Machine Learning}
\begin{figure}[h!]
	\centering
	\begin{tikzpicture}
	
\coordinate (L0) at (0,0);
	
	\draw[line width = 4pt] ($(L0)+(0,0)$)--++(7.1,0);
	
	\draw[line width = 1pt] ($(L0)+(0,0)$)--++(0,-1);
	\draw[line width = 1pt] ($(L0)+(1,0)$)--++(0,-1);
	\draw[line width = 1pt] ($(L0)+(2,0)$)--++(0,-1);
	\draw[line width = 1pt] ($(L0)+(3,0)$)--++(0,-1);
	\draw[line width = 1pt] ($(L0)+(4,0)$)--++(0,-1);
	\draw[line width = 1pt] ($(L0)+(5,0)$)--++(0,-1);
	\draw[line width = 1pt] ($(L0)+(6,0)$)--++(0,-1);
	\draw[line width = 2pt] ($(L0)+(7,0.1)$)--++(0,-1.1);
	
	\MPSiso{$(L0)+(0,0)$}{\norm}{l}{d};
	\MPSiso{$(L0)+(1,0)$}{\norm}{l}{d};
	\MPSiso{$(L0)+(2,0)$}{\norm}{l}{d};
	\MPSiso{$(L0)+(3,0)$}{\norm}{l}{d};
	\MPSiso{$(L0)+(4,0)$}{\norm}{l}{d};
	\MPSiso{$(L0)+(5,0)$}{\norm}{l}{d};
	\MPSiso{$(L0)+(6,0)$}{\norm}{l}{d};
	\MPSiso{$(L0)+(7,0)$}{\norm}{l}{d};

	\draw[line width = 1pt] ($(L0)+(0,-1.2)$)--++(-.15,-.35);
	\draw[line width = 1pt] ($(L0)+(1,-1.2)$)--++(-.15,-.35);
	\draw[line width = 1pt] ($(L0)+(2,-1.2)$)--++(-.15,-.35);
	\draw[line width = 1pt] ($(L0)+(3,-1.2)$)--++(-.15,-.35);
	\draw[line width = 1pt] ($(L0)+(4,-1.2)$)--++(-.15,-.35);
	\draw[line width = 1pt] ($(L0)+(5,-1.2)$)--++(-.15,-.35);
	\draw[line width = 1pt] ($(L0)+(6,-1.2)$)--++(-.15,-.35);
	\draw[line width = 1pt] ($(L0)+(7,-1.2)$)--++(-.15,-.35);
	\node at ($(L0)+(-.2,-1.7)$) {\small $i$};
	\node at ($(L0)+(1-.2,-1.7)$) {\small $i$};
	\node at ($(L0)+(2-.2,-1.7)$) {\small $i$};
	\node at ($(L0)+(3-.2,-1.7)$) {\small $i$};
	\node at ($(L0)+(4-.2,-1.7)$) {\small $i$};
	\node at ($(L0)+(5-.2,-1.7)$) {\small $i$};
	\node at ($(L0)+(6-.2,-1.7)$) {\small $i$};
	\node at ($(L0)+(7-.2,-1.7)$) {\small $i$};
	
	\TNode{}{$(L0)+(0,-1.2)$}{\conj}{};
	\TNode{}{$(L0)+(1,-1.2)$}{\conj}{};
	\TNode{}{$(L0)+(2,-1.2)$}{\conj}{};
	\TNode{}{$(L0)+(3,-1.2)$}{\conj}{};
	\TNode{}{$(L0)+(4,-1.2)$}{\conj}{};
	\TNode{}{$(L0)+(5,-1.2)$}{\conj}{};
	\TNode{}{$(L0)+(6,-1.2)$}{\conj}{};
	\TNode{}{$(L0)+(7,-1.2)$}{\data}{};
	
	\draw[line width = 1pt,rounded corners=.2cm,densely dotted]  ($(L0)+(-.5,-.3)$) rectangle ($(L0)+(7.5,.5)$);
	\draw[line width = 1pt,rounded corners=.2cm,densely dotted]  ($(L0)+(-.5,-1.9)$) rectangle ($(L0)+(6.4,-.8)$);
	\draw[line width = 1pt,rounded corners=.2cm,densely dotted]  ($(L0)+(6.6,-1.9)$) rectangle ($(L0)+(7.4,-.8)$);
	
	\draw[line width= 1pt,densely dotted] ($(L0)+(3.5,.5)$) --++(0,.25);
	\draw[line width= 1pt,densely dotted] ($(L0)+(3,-1.9)$) --++(0,-.25);
	\draw[line width= 1pt,densely dotted] ($(L0)+(7,-1.9)$) --++(0,-.25);
	
	\node at ($(L0)+(3.5,1)$) {MPS $\ket{\Psi}_{DB}$};
	\node at ($(L0)+(3,-2.35)$) {data $\ket{d_i}_D$};
	\node at ($(L0)+(7,-2.35)$) {label $\ket{l_i}_B$};
	
\end{tikzpicture}
	\caption{Architecture of the underlying tensor network. The trainable part of the setup is the MPS, here in blue. The encoded input data of datapoint $i$ of the dataset is represented by the orange tensors. The yellow tensor represents the single site which serves as the output of the label.}\label{Fig:TN architecture}
\end{figure}
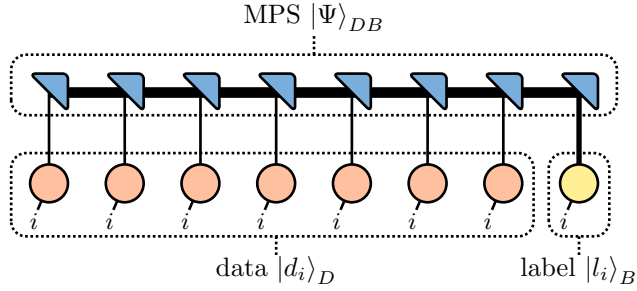

Our tensor network approach is based on an MPS similar to the method outlined in \cite{Stoudenmire2016}. Additionally, we introduce a normalization condition, such that contracting the tensor network yields a normalized vector, see Fig.~\ref{Fig:TN architecture}. Inference of this setup can be run directly on a quantum computer: Either checking each label $\ket{l_i}$ separately or maximally entangling system $B$ with a readout site. The generating MPS can be applied as a quantum circuit following a translation method, e.g. the one given by \cite{Ran2020}.
\par
To maintain a normalized MPS, we use site canonization while sweeping, which is the process of optimizing the sites iteratively until the MPS converges \cite{Ranbook2020}. This approach is found both in gradient based learning \cite{Stoudenmire2016} and in DMRG \cite{Schollwöck2005}. To deal with complex quantum states, we modify the mean square error loss function from \cite{Stoudenmire2016} to be
\par
\begin{equation}\label{eq:loss function}
	L = \frac{1}{2N}\sum_{i=1}^N \big(\bra{f_i}_B - \bra{l_i}_B\big)\big(\ket{f_i}_B-\ket{l_i}_B\big),
\end{equation}
where
\begin{equation}\label{eq:output}
	\ket{f_i}_B = (\bra{d_i}_D\otimes \mathbb{1}_B)\ket{\Psi}_{DB}.
\end{equation}
Here, $\ket{l_i}_B$, the MPS $\ket{\Psi}_{DB}$, andthe data representation $\ket{d_i}_D$ are normalized, but the ML output $\ket{f_i}_B$ is not. Given Eq.~\eqref{eq:output}, we obtain the loss
\begin{equation*}
	L = \frac{1}{2} + \frac{1}{2N} \sum_{i=1}^N \bigg[\bra{\Psi}(\ketbra{d_i}_{D}\otimes\mathbb{1}_B)\ket{\Psi} - 2 \text{Re}\big[\bra{\Psi}_{DB}(\ket{d_i}_D\otimes\ket{l_i}_B)\big]\bigg].
\end{equation*}
As abbrevations, we define the Hermitian $A$ and the vector $b$
\begin{equation*}
	A =  \sum_{i=1}^N  \ketbra{d_i}_{D}\otimes\mathbb{1}_B \quad \text{and} \quad b = \sum_{i=1}^N \ket{d_i}_D\otimes\ket{l_i}_B.
\end{equation*}
For local optimization, one needs locally effective variants $A_s$ and $b_s$ on the site $s$. They are obtained by removing the tensor of the respective site in the MPS and contracting it with $A$ or $b$, \cite{Stoudenmire2016,Schollwöck2005}. Both $A$ and $A_s$ are positive semidefinite. $\braket{k}{b}=0$ if $\ket{k}$ is in the kernel of $A$, otherwise a negative loss could be found, which contradicts Eq.~\eqref{eq:loss function}. Thus, from now on the kernel of $A$ is removed from $A$, when calculating $A^{-1}\ket{b}$. We use the same initialization procedure as \cite{Stoudenmire2016} for the normalized MPS to avoid vanishing gradients. Furthermore, a lookup table of major building blocks of $A_s$ and $b_s$ is maintained while sweeping. This ensures that the sweep only depends linearly on the number of sites, but at the cost of storage space.
\section{Gradient descent with Normalization}
The local optimization problem we tackle is given by
\begin{equation}\label{eq:effective loss function}
    \ket{\Psi_s} = \underset{\ket{\Psi} \,\forall\, \braket{\Psi} = 1}{\argmin}\bra{\Psi}A_s\ket{\Psi} - 2 \text{Re}(\braket{\Psi}{b_s}).
\end{equation} 
If $\ket{\Psi_s}$ is not normalized, the optimum is $A^{-1}\ket{b}$, see \cite{Shewchuk1994,Gerard2006}. Normalizing this optimal solution directly does not yield the optimal solution for the problem with the normalization constraint. This can be seen from the following example:
\begin{equation*}
    A_s = \begin{pmatrix} 2 & 1 \\ 1 & 4 \end{pmatrix} \, \text{and} \, \ket{b_s} = \begin{pmatrix}1 \\ 3\end{pmatrix},\, \text{with} \, \ket{\Psi_s} = \begin{pmatrix}0\\ 1\end{pmatrix} \, \text{but} \, A_s^{-1}\ket{b_s} \approx \begin{pmatrix}0.14\\ 0.71\end{pmatrix}
\end{equation*}
A simple approach to solve the constraint problem is to use the gradient $\ket{g} = A_s\ket{\Psi} - \ket{b_s}$ to give a small update with step size $\alpha$ on $\ket{\Psi}$ and then normalize afterwards. This is known as projected gradient descent \cite{Bubeck2015}. Thus, we obtain 
\begin{equation*}
    \ket{\Psi '} = \frac{\ket{n}}{\sqrt{\braket{n}}} = \frac{\ket{\Psi} - \alpha \ket{g}}{\sqrt{(\bra{\Psi} - \alpha \bra{g})(\ket{\Psi} - \alpha \ket{g})}}.
\end{equation*}
Thus, $\ket{\Psi}$ is updated along the negative gradient on the surface of the unit sphere of $\ket{\Psi}$'s Hilbert space. One simple improvement of the approach is optimizing the step size
$\alpha$. Setting the derivative of the loss w.r.t. $\alpha$ equal to zero yields the condition
\begin{equation}\label{eq:normed gd condition}
    0 =  \text{Re}(\braket{g'}{n})\text{Re}(\braket{n}{g}) - \text{Re}(\braket{g'}{g})\braket{n}
\end{equation}
with the gradient $\ket{g'}$ of $\ket{\Psi'}$ obtained by the next iterative step. We disregard the solution for $\braket{n}\to\infty$ as this is associated with an infinite step size $\alpha$. Calculating $\ket{g'}$ requires one additional matrix-vector multiplication $A_s\ket{g}$ that can be reused to calculate $\ket{\Psi'}$ after the optimal $\alpha$ is found.
\par 
We obtain the derivative of Eq.~\eqref{eq:normed gd condition} by using the product rule and the derivatives:
\begin{equation*}
    \frac{d}{d\alpha} \ket{g} = 0,\, \frac{d}{d\alpha}\ket{n} = - \ket{g},\, \text{and}\,\, \frac{d}{d\alpha}\ket{g'} = A_s\bigg(\frac{\ket{n}\text{Re}(\braket{n}{g})}{\braket{n}^{\frac{3}{2}}}-\frac{\ket{g}}{\sqrt{\braket{n}}}\bigg).
\end{equation*}
We then solve for $\alpha$ with Newton iterations. A local maximum $\alpha \neq \pm \infty$ may exist. To avoid it, $\alpha = 0$ is chosen as the starting point of the Newton iteration.
\section{A Modified DMRG Algorithm}
Optimization techniques based on Krylov subspaces have shown to be far more effective than simple gradient descent \cite{Gerard2006}. Exemplary methods are conjugate gradient descent algorithm, and the Lanczos algorithm used in DMRG. However, DMRG with the underlying Lanczos algorithm does not involve the second term in Eq.~\eqref{eq:effective loss function}, which appears in the machine learning problem. Thus, to modify DMRG to fit this problem, we define a post-processing step after the Lanczos algorithm. First, we compress $A_s$ into $A'_s=VTV^\dag$ with the unitary $V$ and tridiagonal and real $T$ obtained from the Lanczos algorithm. With the fast diagonalization of $T$ \cite{Gerard2006}, we obtain the eigendecomposition of $A'_s = \sum_{i=1}^m e_i \ket{\lambda_i}\bra{\lambda_i}$. The starting vector of the Lanczos iteration is chosen to be $\ket{b_s}$ so that $\ket{b_s}$ is fully represented in the basis of the Lanczos vectors and thus the eigenbasis of $A'_s$.\footnote{Similar lines of reasoning can be found in randomized numerical linear algebra, see \cite{Martinsson2020}.} After compressing, our local effective loss function results in the gradient
\begin{equation}\label{eq:gradient and x}
	 x \ket{\Psi_s} \overset{!}{=} \ket{\text{g}} := A'_s\ket{\Psi_s} - \ket{b_s}.
\end{equation}
For the optimal normalized $\ket{\Psi_s}$ the gradient is not $0$ but equal to $x \ket{\Psi_s}$, with a real scaling factor $x=\pm\sqrt{\braket{\text{g}}}$. This ensures the constraint of the local optimum to the unit sphere on which $\ket{\Psi_s}$ lives, i.e., the gradient is perpendicular to the unit sphere. Eq.~\eqref{eq:gradient and x} can also be derived by Lagrangian optimization with the normalization constraint on $\ket{\Psi_s}$. Assuming $(A'_s-x\mathbb{1})$ is invertible, $\ket{\Psi_s}$ is given by
\begin{equation*}
    \ket{\Psi_s} = (A'_s-x\mathbb{1})^{-1}\ket{b_s}.
\end{equation*}
Given that $\ket{b_s}$ is completely decomposable into the $\ket{\lambda_i}$, we obtain
\begin{equation*}
	\ket{\Psi_s} = ( \sum_{i=1}^N \frac{1}{e_i-x}\ketbra{\lambda_i})\ket{b_s}.
\end{equation*}
We define $b_{si} = \braket{b_s}{\lambda_i} \braket{\lambda_i}{b_s}$, the overlap of the vector $\ket{b_s}$ and the respective eigenvector. With the normalization condition $\braket{\Psi_s} = 1$ we obtain
\begin{equation}\label{eq:equation for x}
	1 =  \sum_{i=1}^m \frac{b_{si}}{(e_i-x)^2}.
\end{equation}
We recall that $A_s$ and subsequently $A'_s$ are positive semidefinite. Thus, their smallest eigenvalue is non-negative $0 \leq e_\text{min}$. Eq.~\eqref{eq:equation for x} suggests multiple solutions $x_s$, yet there is only one $x_s < e_\text{min}$ because the right-hand side of Eq.~\eqref{eq:equation for x} is strictly monotone for $x<e_\text{min}$, i.e. $x_1<x_2 \Leftrightarrow \text{RHS}(x_1) < \text{RHS}(x_2)$.
\par 
Using the compressed $A'$ in the loss function Eq.~\eqref{eq:effective loss function}, we obtain the loss
\begin{equation*}
	L(x) = \sum_{i=1}^m \frac{e_i b_{si}}{(e_i-x)^2} - 2 \frac{b_{si}}{(e_i-x)}
\end{equation*}
with $0<e_i,b_{si}$. For $L(x)$ we find $L(-|x|) \leq L(|x|)$. 
Thus, the global minimum can only be obtained by selecting $x_s<0$. As established above, there can be only one solution $x_s < 0$, which thus is the global minimum. In the gradient descent algorithm, we expect the gradient to converge on the same $\ket{\Psi_s}$. The solution given by $x_s<0$ is obtained because the resulting gradient update is positive and thus amplifies the direction $\ket{\Psi_s}$ over other directions. Thus, the gradient descent algorithm converges on the global minimum. 
\par
Without an analytical solution to Eq.~\eqref{eq:equation for x}, we find this $x_s <0$ by using a Newton iteration. Due to the asymptotical nature of Eq.~\eqref{eq:equation for x}, it only 
converges if the iteration is initialized with $x_s < x_\text{init} < 0$. We find $x_\text{init}$ by checking whether the right-hand side of Eq.~\eqref{eq:equation for x} is larger than $1$. As a side note, during the Lanczos iteration, to avoid numerical instability, we substract $a_i \ketbra{v_k}$ and $\eta_i(\ketbra{v_{k-1}}{v_k}+\ketbra{v_k}{v_{k-1}})$ from $A$, with entries of the tridiagonal matrix $a_i$ and $\eta_i$, and the Lanczos vectors $\ket{v_k}$.
\section{Experimental validation}
{\begin{figure}[h!]
	\centering
	\includegraphics[width=.5\linewidth]{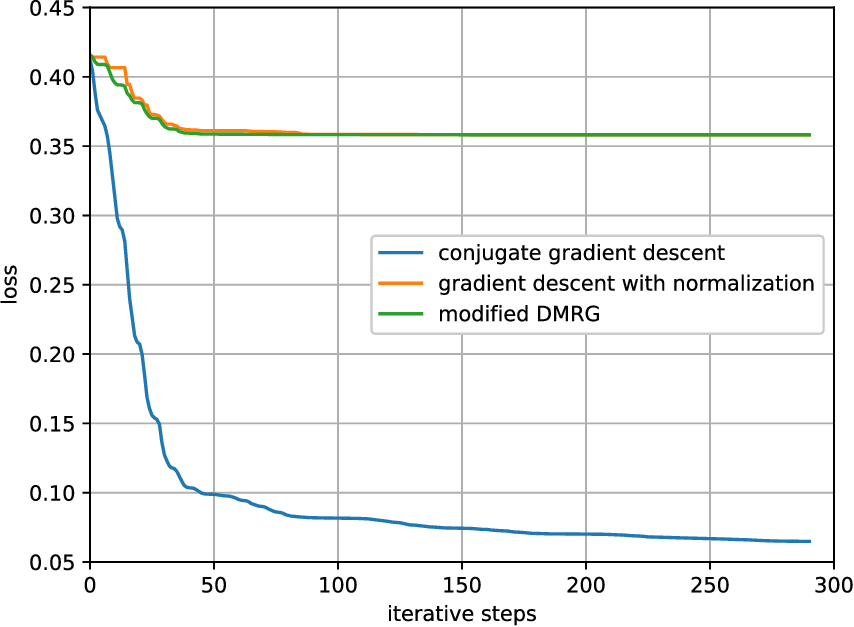}
	\captionsetup{belowskip=0pt}
	\captionsetup{aboveskip=0pt}
	\caption{The loss is displayed over the iterative steps that are given by three sweeps. The modified DMRG method, and the gradient descent with normalization are compared with conjugate gradient descent.}\label{Fig:loss over iterations}
\end{figure}}
To validate our approach and to compare to standard methods, we apply the standard conjugate gradient descent and our two outlined algorithms to the same subset of 5000 MNIST images that are rescaled to 7x7. 4000 images are used in training and 1000 for testing because the matrix $A$ scales with the squar of the number of images. The three approaches were tested for three full two-site sweeps with a maximum bond dimension of 20. Thus, we have $(49\times2 - 1)\times3 = 291$ total local iteration steps, where each iterative step solves our local optimization problem. Fig.~\ref{Fig:loss over iterations} displays the loss over the number of local iteration steps during the sweep. The results for loss and accuracy are given in Tab.~\ref{tab:accuracy}. We observe that introducing the normalization condition significantly impacts both loss and accuracy. However, the norm of the MPS obtained with conjugate gradient descent is $3.9\cdot 10^6$. Normalizing this MPS, gives only vanishing overlaps in the loss function Eq.~\eqref{eq:loss function}, which results in a trivial loss of $\approx 0.5$. 
{\begin{table}[h!]
    \centering
    \begin{tabular}{c|c|c|c|c}
    method                      &    training loss  & test loss & training acc. & testing acc.  \\\hline
    initialization              &     $0.43646$     & $0.43984$ &   $67.725\%$      &  $63.300\%$       \\
    conjugate grad. desc.       &     $0.06487$     & $0.08680$ &   $96.950\%$      &  $94.700\%$       \\
    norm. grad. desc.           &     $0.35820$     & $0.36256$ &   $75.750\%$      &  $73.000\%$       \\
    modified Lanczos            &     $0.35820$     & $0.36257$ &   $75.775\%$      &  $73.100\%$       
    \end{tabular}
	\captionsetup{belowskip=-15pt}
    \caption{Overview of accuracy and loss of the different methods.}
    \label{tab:accuracy}
\end{table}}
\section{Conclusion and Outlook}

In this work, we take the step to adapt tensor network methods to approach quantum machine learning more efficiently. Implementing the normalization condition is necessary to make the algorithm fit for deployment on quantum computers. However, the results of the validation show that the performance needs to be increased to be competitive with other contemporary algorithms. Clearly, further improvements are needed to achieve parity with classical tensor network methods. To this end, more complex normalization conditions should be studied, like those in quantum channels and improvements in the evaluation of the methods need to be developed. Overall, major research still needs to be done to bring machine learning efficiently to quantum computers.

\begin{footnotesize}
\bibliographystyle{unsrt}
\bibliography{literature.bib}
\end{footnotesize}

\end{document}